\documentstyle[psfig]{mn}

\def\jcd{Christensen-Dalsgaard}
\def\note #1]{{\bf #1]}}
\def\alphaas{\alpha_{\rm as}}
\def\alphac{\alpha_{\rm c}}
\def\K{\,{\rm K}}
\def\Teff{T_{\rm eff}}
\def\Xc{X_{\rm c}}
\def\rt{r_{\rm t}}
\def\Yenv{Y_{\rm env}}
\def\Htwo{{\cal H}_2}
\def\Htwof{{\cal H}_2^{\rm f}}
\def\Gf{G^{\rm f}}
\def\Gas{G_{\rm as}}
\def\Gcow{G_{\rm cow}}
\def\Gasf{G_{\rm as}^{\rm f}}
\def\Gfas{G_{\rm as}^{\rm f}}
\def\Gfcow{G_{\rm cow}^{\rm f}}
\def\wa{\omega_{\rm a}}
\def\wc{\omega_{\rm c}}
\def\pp{{\rm p}}
\def\dd{{\rm d}}
\def\muHz{\, \mu{\rm Hz}}

\begin{document}

\title[Phase function for solar-like stars]
{The phase function for stellar acoustic oscillations - IV. 
Solar-like stars}

\author[F. P\'erez Hern\'andez and J. Christensen-Dalsgaard]
{F. P\'erez Hern\'andez$^{1}$\thanks{E-mail: fph@iac.es}
and J. Christensen-Dalsgaard $^{2}$\thanks{E-mail: jcd@obs.aau.dk}\\
$^{1}$Instituto de Astrof\'{\i}sica de Canarias, E-38200 La Laguna,
Tenerife, Spain\\
$^{2}$Teoretisk Astrofysik Center, Danmarks Grundforskningsfond and
Institut for Fysik og Astronomi, Aarhus Universitet, \\
DK-8000 Aarhus C, Denmark}

\date{Accepted ????. 
Received ????; 
In original form 1997 June ??}

\pagerange{\pageref{firstpage}---\pageref{lastpage}}
\pubyear{????}

\label{firstpage}

\maketitle
\begin{abstract}
In recent years there has been some progress
towards detecting solar-like
oscillations in stars. The goal of this challenging project
is to analyse frequency spectra similar to that observed for the Sun in
integrated light.
In this context it is important to investigate what can 
be learned about the structure and evolution of the stars from such future
observations. Here we concentrate 
on the structure of the
upper layers, as reflected in the phase function.
We show that it is possible to obtain this function from low-degree
$\pp$ modes, at least for stars on the main sequence.
We analyse its dependence on several uncertainties in the structure 
of the uppermost layers.
We also investigate a filtered phase function, which has
properties that depend on the layers around the second helium 
ionization zone.
\end{abstract}

\section{Introduction}

Since the successful development of helioseismology in the last few decades,
the interest in extending the work to other stars has become clear.
However, the small amplitudes expected in the power spectra of stars
with solar-like oscillations makes this a very difficult task and to
date no unambiguous detection has been made.
The main limitation is the atmospheric noise. 

On the other hand, these stellar spectra are expected to have regular
patterns that can be detected even without a full determination of the
$\pp$-mode frequencies. Some efforts have been made in this direction 
(Brown et al.\ 1991; Pottasch, Butcher \& van Hoesel 1992;
Kjeldsen et al. 1995).
The most interesting parameters that can be obtained in this 
limit are the so-called small and large separations,
as demonstrated by several theoretical analyses 
(e.g.\ \jcd \ 1984, 1988, 1993; Ulrich 1986;
Gough 1987; Gough \& Novotny 1993).

However, the full power of
asteroseismology for solar-like stars would, as in the case of
helioseismology, require extensive determination of 
individual $\pp$-mode frequencies.
In fact, once the observed amplitudes are above the noise
level, a few weeks of observations will be enough to obtain accurate
frequency measurements.
In the near future there
will be space missions devoted to asteroseismology (Baglin 1991; 
Catala et al.\ 1995)
which, we hope, will give the first firm results on this issue. 
With these future observations in mind,
in this paper we shall assume that a full set of low-degree ($l\leq 2$)
$\pp$-mode frequencies are available
in a given frequency range. We do not consider modes of higher degree because
they cannot be detected with simple photometric observations.
Also, we suppose that the radial order $n$ and the degree $l$ of the modes are 
known, and, of course, that the frequencies have been corrected for their 
rotational splittings. The determination of $n$ and $l$ requires 
comparison with the models but for solar-like stars on the main sequence,
such as we shall consider here, it is plausible to assume that this 
is possible without any uncertainty.

As in the solar case, a direct comparison of theoretical and observed 
frequencies would be inconclusive. 
However, stellar acoustic oscillations satisfy a simple 
asymptotic relation which allows the separation of the contribution 
of the upper layers, such as the convective envelope, from that of the deep
interior. 
This relation has been used extensively in the solar case, 
for instance to obtain the sound speed (e.g.\ \jcd , Gough \& Thompson 1989) 
and to analyse 
the upper layers (e.g.\ P\'erez Hern\'andez \& \jcd \ 1994b).
Since for the distant stars 
we expect to detect only low-degree modes, 
we re-analyse the relation in this context. 
A similar investigation was carried out recently by Lopes et al. (1997),
to investigate what would be the diagnostic potential of observing solar-like
oscillations in the star $\beta$ Virginis.

Analysis of observations 
of low-degree solar $\pp$-modes is illustrative of what can
be inferred for other stars.
Thus we first present an analysis of the observations by
Lazrek et al. (1997), obtained with the GOLF instrument on
the SOHO spacecraft.
Having demonstrated that it is possible to carry out the asymptotic analysis
for stars,  
we concentrate our work on the upper layers as probed by the so-called 
phase function. We investigate the information that can be obtained from this
function, 
taking into account the plausible accuracy of the frequency determinations.
We pay particular attention to the second helium ionization zone. 

\section{The Sun as a star}

\subsection{The phase function for low-degree modes}

Neglecting the perturbations in the gravitational potential and considering
the waves locally as plano-parallel under constant gravity, the
frequencies of $\pp$ modes satisfy the asymptotic relation \cite{gough}
\begin{equation}
\frac{ \pi (n+\epsilon ) }{\omega_{nl} } \simeq  
\int_{r_1}^{r_2} \left[ 1 - \frac{\wc^2}{\omega^2}- \frac{S_l^2}
{\omega^2} \left(1-\frac{N^2}{\omega^2}\right)\right]^{1/2}  \frac{\dd r}{c} 
\; .
\label{as1}
\end{equation}
Here $\omega$ is the angular frequency, $c$ 
the adiabatic sound speed, $N$ the buoyancy frequency, 
$\wc$ the cut-off frequency and $S_l$ the Lamb acoustic frequency.
The integral is over radius $r$ with the limits $r_1$
and $r_2$ defined by the vanishing of the bracket in the integral.
In principle, the asymptotic theory gives
$\epsilon = -1/2$ but since close to the surface the asymptotic conditions
are not satisfied, $\epsilon$ is assumed to be a function of frequency,
yet to be determined
-- for low-degree modes its dependence on $l$ is negligible. 

It is interesting to note that with the sole assumption that the perturbations
in the gravitational potential can be neglected, equation~(\ref{as1}) 
still holds in the asymptotic limit, 
provided $\wc$ and $N$ are replaced by more general
functions \cite{gough2}. 

For the Sun, and many other stars, equation~(\ref{as1}) 
can be further approximated
to yield a very useful relation.
If there is a point $r_0$ -- usually in the convective envelope -- 
such that $|N^2 (r)|\ll \omega^2$ and 
$\wc^2 (r)\ll \omega^2$ for $r<r_0$ and  
$S_l^2 \ll \omega^2$ for $r>r_0$, then it can be shown that 
equation~(\ref{as1}) 
can be approximated by (e.g.\ \jcd \ \& P\'erez Hern\'andez 1992)

\begin{equation}
\frac{ \pi n }{\omega_{nl} } \simeq  F(\omega /L) - G(\omega )
\; ,
\label{as2}
\end{equation}
where
\begin{equation}
F(w) = 
\int_{\rt}^{R} \left( 1 - \frac{c^2}
{w^2 r^2} \right)^{1/2}  \frac{\dd r}{c} 
\label{fw}
\end{equation}
and
\begin{equation}
G(\omega ) = \frac{\pi\alpha (\omega )}{\omega }
\; .
\label{gw}
\end{equation}
Here $L = l + 1/2$, $l$ being the degree of the mode, and $w = \omega / L$.
The integral in equation~(\ref{fw}) 
is from the lower turning point
$\rt$ (defined by the vanishing of the bracket in the integral)
to the surface radius $R$. Finally,
the phase function $\alpha(\omega)$ depends on conditions
near the stellar surface.

\begin{figure}
 \centerline{\psfig{figure=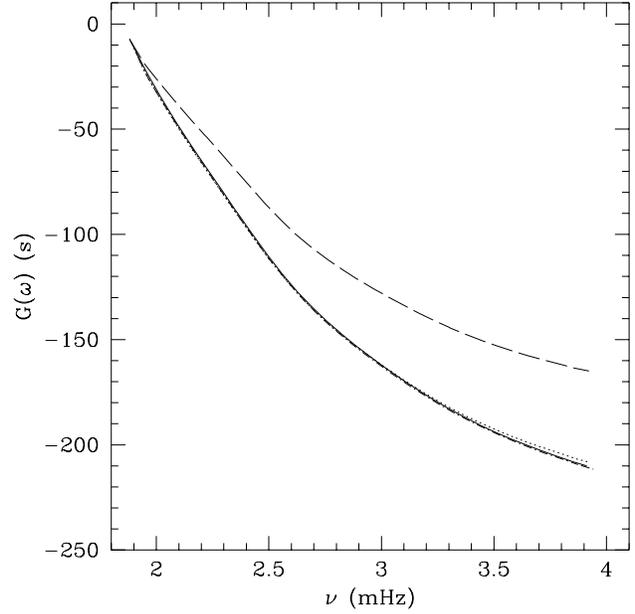,width=.49\textwidth}}
 \caption{
The continuous line is $\Gas (\omega )$ for Model~A. 
The dashed line is $G(\omega )$ for the same model and the frequency set 
indicated in the text. 
The dot-dashed line is $\Gcow(\omega )$ for the same model and mode-frequency 
set but in the Cowling approximation. 
The dotted line is $G +d/\omega ^2$, where $d$ is a constant. 
The functions have been shifted by constants to
match at the lowest frequency.}
 \label{fig1}
\end{figure}

This phase function can be computed 
from the structure of the upper layers of a stellar
model, by fitting numerically computed eigenfunctions to an asymptotic
approximation at a point where the latter is valid
(see \jcd \ \& P\'erez Hern\'andez 1992). We 
denote the phase function obtained in this way $\alphaas (\omega )$
and the corresponding related function as defined in
equation~(\ref{gw}) is denoted $\Gas (\omega )$. 
Since, in this case, the wave equations 
are integrated from the surface down to a point in the envelope,
by construction $\alphaas$ and $\Gas$ depend only on the upper layers.
The continuous line in
Fig.~\ref{fig1} corresponds to $\Gas(\omega )$ 
for a solar model without diffusion,
(\jcd , Proffitt \& Thompson 1993), in the following 
referred to as Model~A.

It is interesting to note that the well-known Tassoul equation 
\cite{tassoul},
which is an approximation valid for low-degree modes,
\begin{equation}
\nu_{nl}  \simeq \left( n+ \frac{L}{2} + \theta \right ) \Delta\nu
- \left (A L^2 -
\delta \right ) \frac{\Delta\nu^2}{\nu_{nl}}
\; ,
\label{tas}
\end{equation}
where $\nu_{nl}$ is the cyclic frequency and $\theta$, $\delta$, $A$ and  
$\Delta\nu$ are constants, is a particular
case of equation~(\ref{as2}). 
In fact, it is straightforward to show that in this 
approximation
\begin{equation}
F (w)   \simeq \frac{1}{2\Delta\nu} - \frac{\pi}{2w} 
+ \frac{2\pi^2 A \Delta\nu}{w^2}\label{fwtas}
\label{lowlf}
\end{equation}
and
\begin{equation}
G(\omega )  \simeq \frac{\pi\theta}{\omega } + \frac {2\pi^2\delta \Delta\nu}
{\omega^2} \; .
\end{equation}
A detailed derivation of equation~(\ref{tas}) from equation~(\ref{as2}) 
was given by Vorontsov (1991)
(see also Gough 1986).

If the radial order $n$ and the degree $l$ 
are assumed to be known, $F(w)$ and $G (\omega )$
can be estimated from $\pp$-mode frequencies by fitting 
them to equation~(\ref{as2}). 
Hence these functions are
potentially observable quantities for solar-like stars.
This is the advantage of using equation~(\ref{as2}) rather 
than the formally more precise
equation~(\ref{as1}).
In the present work
we have made a least-squares fit, expanding
$F(w)$ and $G(\omega )$ in terms of Legendre polynomials
in $w$ and $\omega$:
\begin{equation}
\frac{ \pi n }{\omega_{nl} }  \simeq a_0+\sum_{i=1}^k a_i f_i (x)
+ \sum_{i=1}^m b_i g_i (y)
\; ,
\label{fitfun}
\end{equation}
where $f_i$ and $g_i$ are Legendre polynomials of order $i$ with coefficients
$a_i$ and $b_i$. The variables 
$x$ and $y$ are linearly related to $w$ and $\omega$, respectively, 
and are defined in the interval $[-1,1]$.
Note that from such a fit $F(w)$ and $G (\omega )$ can be obtained 
only to within a constant. 

Throughout the paper we use the same mode data set, consisting of 
modes with $l =0,1,2$ and $12\leq n \leq 27 $. The interval in $n$
has been chosen such as to reject modes with very low
amplitudes, which are not expected to be observed in solar-like stars.
We note that rather similar results are obtained if modes with $l=3$ 
are added.
The fit was carried out through $\chi^2$ minimization,
using frequency errors obtained in early observations
with the GOLF instrument on the SoHO spacecraft (Lazrek et al. 1997).
In equation~(\ref{fitfun})
we have used Legendre polynomials of order 14 for each function.
We note in passing that,
as suggested by equation (\ref{lowlf}), the expansion
of $F(w)$ is most reasonably carried out in terms of $w^{-1}$.
In this case a value $k\simeq 5$ 
in the expansion of $F(w)$ is sufficient to achieve a good fit.
But in any case this
does not affect to the determination of $G(\omega )$ with which we are mainly
concerned in this work.

This kind of fit for $G(\omega )$ is rather similar to that used in 
inversion techniques (e.g.\ Dziembowski, Pamyatnykh \& Sienkiewicz 1990) 
though in that case the fit is usually done
to a function corresponding to differences between models 
rather than to $G(\omega )$ directly.
On the other hand, Vorontsov, Baturin \& Pamyatnykh (1992)
applied a very similar fit to solar $\pp$ modes,
but considering modes with higher degrees.

For the Sun, equation~(\ref{as2}) 
is a good approximation for modes of moderate degree,
in particular for those modes with inner turning points between the base 
of the convection zone and the second helium ionization zone.
When this kind of mode set is used, the asymptotic
function $\Gas (\omega )$ and the numerically fitted $G(\omega )$ 
agree very well. Therefore,
the observational $G(\omega )$ can be used as a test of the upper layers.
However, 
when only low-degree modes are included, such as is the case for other stars, 
the approximations leading to equation~(\ref{as2}) are not 
so accurate and hence larger differences between $\Gas$ and $G$ are
expected.
This can be seen in Fig.~\ref{fig1}, where the dashed line 
corresponds to $G(\omega )$ for Model~A (since $G$ is obtained to within a
constant, in Fig.~\ref{fig1} 
and the following figures we choose the constant
such that all the functions agree at a given frequency).
Indeed, there are significant differences between $G(\omega )$ and 
$\Gas(\omega )$,
the main difference coming from neglecting the perturbations in 
the gravitational potential in equation~(\ref{as2}).
This can easily be checked by 
determining the corresponding function
$\Gcow(\omega)$ by fitting to $\pp$-mode frequencies 
computed in the Cowling approximation.
This is shown by the dot-dashed line in Fig.~\ref{fig1}.
It is indeed much closer to $\Gas$. 
Thus, the function
$G$ obtained from the observations is a sum of $\Gas\simeq \Gcow$  
that depends on the upper layers  and a function 
that depends on the perturbation in the gravitational potential
throughout the solar interior.

For completeness we also note that the function $F(w)$ obtained from a fit
to equation~(\ref{as2}) does not
agree with the asymptotic expression~(\ref{fw}), except if the frequencies in
the Cowling approximation are used.

\subsection{Perturbations on the gravitational potential}

In principle, equation~(\ref{as2}) can be improved 
by including the perturbation in the gravitational potential. 
To first order, the result is an additional term 
with a specific dependence on $\omega $ and $L$, 
such that equation~(\ref{as2}) is replaced by
\begin{equation}
\frac{ \pi n }{\omega_{nl} } \simeq  F\left(\frac{\omega}{L}\right) 
- G(\omega ) + \frac{1}{\omega^2}P_{\Phi}\left(\frac{\omega}{L}\right)
\label{as3}
\end{equation}
\cite{voro}.
It is straightforward to show that if $F(w)$, $G(\omega )$ and $P_{\Phi}(w)$
are solutions of a fit to equation~(\ref{as3}) then so are
$F(w) + C$, $G(\omega) + C + D/\omega^2$ and $P_{\Phi}(w) + D$ for any
constants $C$ and $D$. 
In particular, it is not possible to separate the 
contributions of the upper layers, as given by $\Gas$, from the zero-order
term of $P_{\Phi}$ since in any fit to observational data both would be 
included in $G(\omega )$.

We consider again the function $G$ obtained
from a fit to equation~(\ref{as2}).
In Fig.~\ref{fig1} the dotted line represents $G(\omega ) + d/\omega^2 $
with a suitable (positive) value for the constant $d$
(as well as including the constant shift mentioned above).
It is clear that $G \simeq \Gas - d/\omega ^2$. 
Had we done a fit to equation~(\ref{as3}), we would have obtained a similar
discrepancy due to the undetermined constant $D$.
Thus, in what concerns the determination of $G$, 
a fit to equation~(\ref{as3}) is equivalent to a fit to equation~(\ref{as2}).

We shall now argue that, when considering a pair of models (or a model and
the observations), the differences in $P_{\Phi}$ are very small and hence
$\delta G  \simeq \delta \Gas$.
To a first approximation, 
the function $P_{\Phi}$ is given by (e.g.
Vorontsov 1991; Christensen-Dalsgaard 1996)
\begin{equation}
P_{\Phi}\left(\frac{\omega}{L}\right) = 2\pi G
\int_{\rt}^R \rho \left( 1 - \frac{L^2 c^2}
{\omega^2 r^2} \right )^{-1/2} \frac{\dd r}{c}
\; ,
\label{pw}
\end{equation}
where $\rho$ is the density.
Since we are dealing with low-degree modes, $P_{\Phi}$
can be approximated by a Taylor expansion in $\tilde w = L/\omega$ around
the centre. From equation~(\ref{pw}), we have
\begin{equation}
P_{\Phi}(\tilde w = 0) =  2\pi G \int_0^R \rho \frac{\dd r}{c}
\propto \bar\rho^{1/2}\int_0^1 \frac{\tilde\rho}{\tilde c} \dd x
\label{pw0} 
\; ,
\end{equation}
where $\bar\rho$ is the mean density, $x=r/R$ and $\tilde \rho$, 
$\tilde c$ are dimensionless functions.
Also, by setting $z=r^2/c^2$ as the independent variable and 
integrating equation~(\ref{pw})  by parts twice, it follows that
\begin{equation}
\frac{\dd P_{\Phi}}{\dd\tilde w}\Big|_{\tilde w = 0}=0
\; .
\end{equation}
Thus, the variation of $P_{\Phi}$ is of second order in $\tilde w$,
and in the $\tilde w$ interval spanned by 
our data set $P_{\Phi}$ can be estimated by its value at the centre.

{}From equation~(\ref{pw0}), it follows that, 
for models with the same mean density and
similar structure in the inner region, the functions
$P_{\Phi}$ are similar;
hence, when considering differences between 
pair of such models, this term can be neglected.
In fact, numerical tests show
that for realistic solar models the differences between
$G(\omega )$ are very close to the corresponding 
differences in $\Gas (\omega )$.
Hence, although the function $G(\omega )$ 
depends also on properties of the interior, 
differences between such functions depend mainly on the differences 
between the uppermost layers.
For distant stars the mean density is generally not known
with sufficient accuracy;
however, in this case it is likely that similar cancellation
can be achieved by working in terms of frequencies measured
in units of the large frequency separation 
$\Delta \nu_{nl} = \nu_{nl} - \nu_{n-1 \, l}$
(cf. equation \ref{tas}) which scales approximately as $\bar\rho^{-1/2}$.

\begin{figure}
 \centerline{\psfig{figure=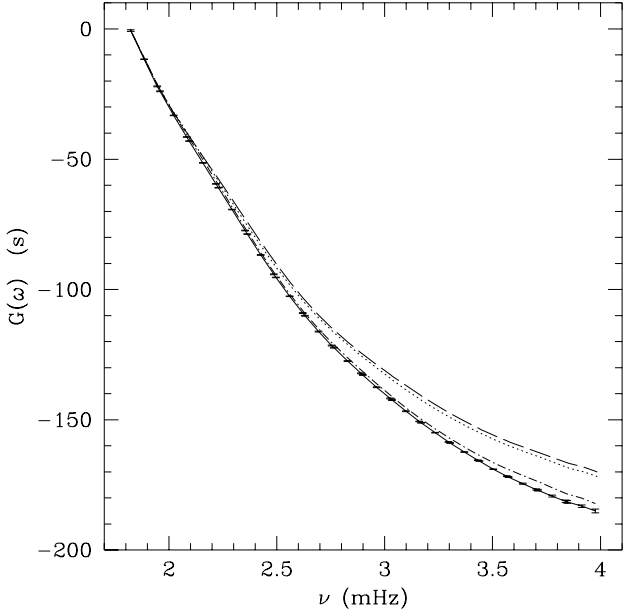,width=.49\textwidth}}
 \caption
{The function $G(\omega )$ for the observations (continuous line
with error bars),
Model~A (dashed line), Model~B (dotted line) and Model~C
(dot-dashed line).
The functions have been shifted by constants to match at the lowest frequency.
}
 \label{fig2}
\end{figure}

In order to illustrate the dependence of $G(\omega )$ on stellar 
structure, in Fig.~\ref{fig2} we show this function for three solar models and
the mode set indicated above.
The dashed line corresponds to the previously mentioned Model~A, 
the dotted line to a model that includes helium settling 
(Christensen-Dalsgaard, Proffitt \& Thompson 1993, hereafter Model~B) 
and the dot-dashed line to a solar model that in
addition has an artificial increase in the surface opacities in order 
to get a better agreement with the observations (hereafter Model~C). 
As a result of helium settling, the envelope helium
abundance $\Yenv \simeq 0.25$ of Models~B and C
is lower than the value $\simeq 0.28$ for Model~A.
On the other hand, Model~C differs
substantially from the other two in the uppermost layers.
As can be seen, the function $G(\omega )$ for this model
is also significantly different.
In Fig.~\ref{fig2} 
we also show $G(\omega )$ computed with the observed frequencies
\cite{golf},
with errors estimated by Monte-Carlo simulation.
As expected, Model~C is closest to the observational data.
(Note that, since the superficial layers of the solar model
suffers from other
uncertainties, such as non-adiabatic and convective effects, 
it cannot on this basis be concluded that Model~C 
is better than the other two.)
This result is similar to that obtained by using intermediate-degree modes. 

For distant stars with 
poorly known parameters,
the uncertainty in $P_{\Phi}(0)/\omega^2$ 
can eventually become comparable with the differences in $\Gas$. 
In this case, it may be better to work with a phase function invariant
under the transformation $G  ' \rightarrow G+C+D/\omega^2$ or, equivalently,
$\alpha  ' \rightarrow \alpha + C\omega + D/ \omega $.
It is easy to show that the phase function 
\begin{equation}
\gamma = \alpha - \omega \frac{\dd\alpha}{\dd\omega} - \omega^2 
\frac{\dd^2\alpha}{\dd\omega^2}
\end{equation}
has this property and hence depends only on the upper layers. 
This is a generalization of 
\begin{equation}
\beta = \alpha - \omega \frac{\dd\alpha}{\dd\omega} \; , 
\end{equation}
which has been used in earlier investigations (e.g. Brodsky \& Vorontsov 1988; 
Christensen-Dalsgaard \& P\'erez-Hern\'andez 1991; Lopes et al. 1997), 
and which is invariant only under
the transformation $\alpha ' \rightarrow \alpha + C\omega $.

We note that the function $\gamma$ introduces a filter 
which suppresses partially the contribution of
the uppermost layers. Thus, if 
a good estimate of the uncertainty in $P_{\Phi}$ as given by 
equation~(\ref{pw0}) is available,
it is generally better to work with $\alpha(\omega )$ or $G(\omega )$. 
On the other hand, if we are interested in deeper layers 
(such as the second helium ionization zone), then
the filtering procedure described below may be preferable,
in isolating more effectively the contribution from just these layers.

\subsection{The filtered phase function}

Although $G$ depends mainly on the uppermost layers, it also has a small
contribution from deeper layers.
In fact, it is possible to obtain from $G(\omega )$ a function which 
predominantly reflects the properties
of the layers around the second helium ionization zone.

P\'erez Hern\'andez \& Christensen-Dalsgaard (1994a) 
showed how a similar function can be 
extracted through a filter. Rather than equation~(\ref{as2}), they considered 
an equivalent expression for differences,
assumed to be small, between pairs of models.
With this assumption
the differences between phase functions are linearly 
related to differences between the
equilibrium models \cite{paperi} by
\begin{equation}
\Htwo\equiv \frac{\pi\delta\alpha}{\omega}
\simeq
\int_{r_0}^R\left[ K_c (r )\frac{\delta c}{c}
(r)+ K_{\wa}(r )\frac{\delta\omega_{\rm a}}{\omega_{\rm a}}(r)\right]
{\rm d} r  
\; ,
\label{h2}
\end{equation}
where $\wa$ is the Lamb acoustical cut-off frequency and the 
kernels $K_c$ and $K_{\wa}$ are known functions. 
Due to the behaviour of the kernels in the solar interior, a localized
perturbation has a frequency dependence of the form
(P\'erez Hern\'andez \& Christensen-Dalsgaard, 1994a) 
\begin{equation}
\Htwo \propto \cos  2 \big[\omega \tau - (\alpha +1/4)\pi\big]
\; ,
\label{h2p}
\end{equation}
where $\tau$ is the acoustical depth,
\begin{equation}
\tau = \int_r^R {1 \over c} \dd r
\; .
\label{tau}
\end{equation}
Hence shorter-period components in $\Htwo$ 
correspond to perturbations in deeper layers.
Thus by filtering the smooth components of $\Htwo$ it is possible to
obtain a function $\Htwof$ that depends on deeper layers. 

In the present case, $G(\omega )$ can be thought of as the
difference between the actual model (or the Sun) and a smoothed model that
does not have this localized feature. 
Relation (\ref{h2p}) is then still valid
and we can obtain a filtered $\Gf$ in a similar way to $\Htwof$.

The filter used by P\'erez Hern\'andez \& Christensen-Dalsgaard (1994a) 
was a recursive filter 
especially suitable for
$\Htwo$ because of the flatness of its smooth component 
at low frequencies ($\nu \leq 1.4$ mHz for the Sun). 
In the stellar case modes at such
low frequencies are unlikely to be observed because of their
expected low surface amplitudes. 
Furthermore, for the function $G$, upon which much of
our analysis is based, the smooth component does not have this behaviour.
However, below we describe a different 
filter which has similar properties 
and can be extracted directly from the least-squares fit used to compute 
$G(\omega )$.

Since, in the present work, we obtain $G$ as a Legendre
polynomial expansion, a simple way  of separating the smooth and oscillatory
frequency patterns is to consider the low- and high-order polynomial 
coefficients, respectively.
A very similar type of filtering was introduced by 
Vorontsov et al.\ (1992).
Specifically, we define
\begin{equation}
\Gf (\omega ) = \sum_{i=p+1}^m b_i g_i (\omega )
\; .
\label{gf}
\end{equation}
A suitable value of $p$ will give the desired oscillatory function.
Similarly, the asymptotic function $\Gas$ can be fitted to Legendre polynomials,
from which $\Gasf$ may be defined.

\begin{figure}
 \centerline{\psfig{figure=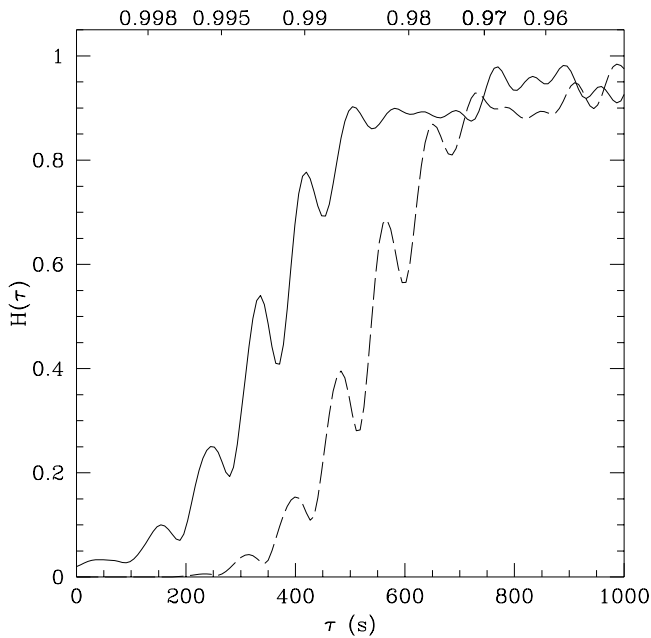,width=.49\textwidth}}
 \caption
{Response function $H(\tau )$ as function of acoustical depth $\tau$
and radius (on top). The solid line is for equation~(\protect\ref{xrf}) 
and the dashed line for harmonic functions,
both for fits with $p = 6$. }
 \label{fig3}
\end{figure}

In order to choose the value of $p$ in equation~(\ref{gf}),
we have computed the response to the filtering of signals
of the form (\ref{h2p}).
Specifically, we define the response function $H(\tau)$ by 
\begin{equation}
H(\tau ) = \frac{\sum_i y_i^2(\omega_i , \tau )}
{\sum_i x_i^2(\omega_i ,\tau )}
\; ,
\label{rf}
\end{equation}
where $x (\omega , \tau)$ is 
the input signal and $y (\omega , \tau)$ is the output function,
calculated as in equation (\ref{gf}) by expanding $x(\omega , \tau)$ 
in Legendre polynomials of order $m$ 
and setting to zero the first $p$ terms;
the summation is over the frequency range considered.
Here we have used the observed frequencies of Lazrek et al. (1997).

We consider filtering corresponding to $p = 6$.
The dashed line in Fig.~\ref{fig3} 
shows the response to simple harmonic functions
$x_i = \cos(2\omega_i \tau)$. 
However, since in equation~(\ref{h2p}) $\alpha$ is a function
of frequency, the contribution to $G(\omega )$ 
of a given layer is not exactly a harmonic function. 
To investigate the effect of this, in equation~(\ref{rf}) we take as input
\begin{equation}
x_i = \cos 2 \big[\omega_i \tau - (\alphaas(\omega_i)+1/4) \pi \big] \; ,
\label{xrf}
\end{equation}
with $\alphaas$ computed for a given solar model.
The solid line in Fig.~\ref{fig3} 
corresponds to $H(\tau )$ calculated in this way, 
with $\alphaas$ obtained from Model~A.
This term clearly has a significant
effect, corresponding approximately to a shift in $\tau$. 
{}From the figure, it follows that the filter 
considered suppresses most of the signal 
from the uppermost layers ($r > 0.997R$) while passing the signal from
$r \sim 0.98R$, where the second helium ionization zone is located.
Hence in the following we shall take $p=6$ in equation~(\ref{gf}), at least
for the Sun.

\begin{figure}
 \centerline{\psfig{figure=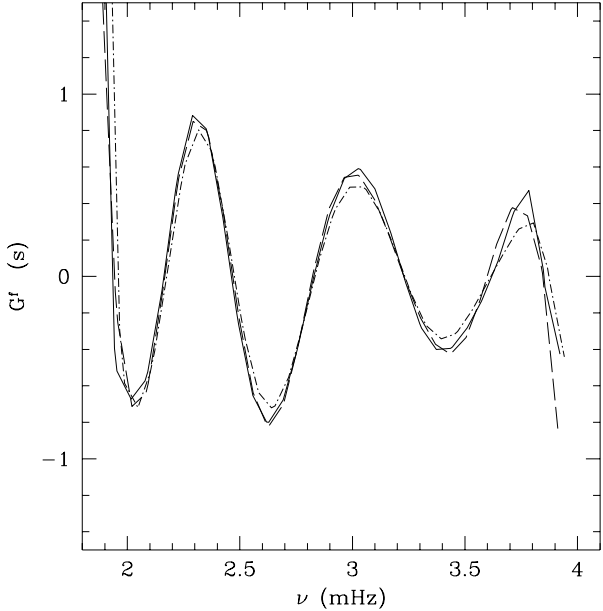,width=.49\textwidth}}
 \caption
{The dashed line is $\Gfas(\omega )$ for Model~A. 
The solid line is $\Gf(\omega )$ for the same model and the dot-dashed
line the same function but using $\pp$-mode frequencies in the Cowling
approximation. }
 \label{fig4}
\end{figure}

In Fig.~\ref{fig4} 
we show $\Gasf (\omega )$ and  $\Gf(\omega )$ for a solar model. 
The dashed line corresponds to $\Gfas$, the dot-dashed
line to $\Gfcow$ computed in the Cowling approximation and the solid line
to $\Gf$ using the full equations. 
Since $\Gf  \simeq \Gfcow$, $\Gf$ does not depend on the perturbations
in the gravitational potential. Furthermore, since $\Gf  \simeq \Gfas$,
the function $\Gf$ depends on the layers between $r_0$ 
(a point in the adiabatically
stratified convection zone) and an upper limit given by the response 
function, as in Fig.~\ref{fig3}.

\begin{figure}
 \centerline{\psfig{figure=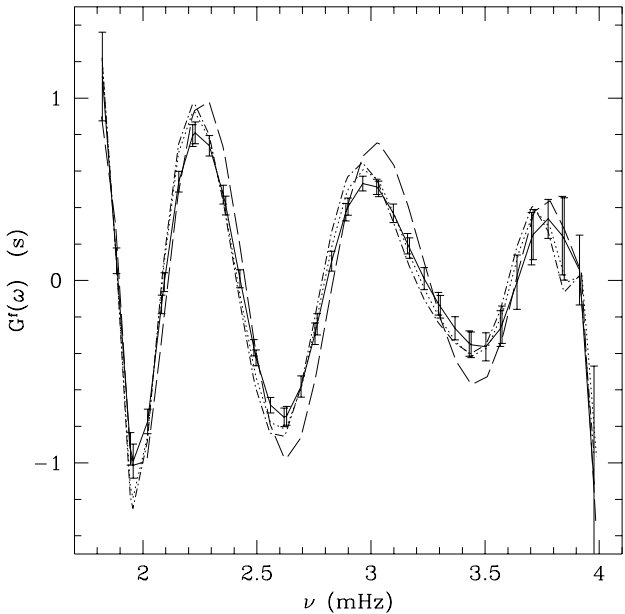,width=.49\textwidth}}
 \caption
{The continuous line with error bars is $\Gf$ for the observations, 
the dashed line is
$\Gf$ for Model~A, the dotted line is for
Model~B, and the dot-dashed line is for Model~C. }
 \label{fig5}
\end{figure}

In Fig.~\ref{fig5} 
we show $\Gf (\omega )$ for the same solar models as in Fig.~\ref{fig2}.
The function $\Gf$ is very similar for Models~B and C,
showing that the filtering has suppressed
the substantial difference in $G$ between Models B and C
(cf. Fig.~\ref{fig2}), caused by the differences
in the structure of the superficial layers.
On the other hand $\Gf$ for Model~A is clearly different,
reflecting the differences in
the equilibrium structure in the second helium ionization zone. 
As shown by P\'erez Hern\'andez and \jcd \ (1994b)
this kind of filtered phase function depends on three main properties:
the equation
of state, the helium abundance $\Yenv$ in the envelope 
and the specific entropy $s$ in the adiabatically stratified part 
of the convection zone, whose value controls the depth of the 
convection zone.
Model~A has the same equation of state than Models
B and C but $\Yenv$ is larger and the depth of the convection zone
smaller. 
In Fig.~\ref{fig5}
we also show $\Gf$ for the observations \cite{golf},
with errors determined from a Monte-Carlo simulation.
It can be seen that the models with diffusion (and hence $\Yenv\simeq 0.25$
and a depth of the convection zone closer to that inferred from 
sound speed inversions)
agree better with the observations.
This is in accordance with the the analysis with moderate-degree modes, as
shown in P\'erez Hern\'andez \& Christensen-Dalsgaard (1994b). 
Although in the present case the errors are larger,
our purpose is to show
that by using low-degree modes alone it is possible to carry out a similar
analysis for the distant stars. 

\section{Phase function for stars}

\subsection{Global properties}

In this section we shall consider the phase functions $G(\omega )$ 
and $\Gf(\omega )$ for models of stars on the main sequence. 
The models considered are summarized in Table~\ref{mod}.
There are two models for each mass, one at the 
zero-age main sequence and the other near the
end of hydrogen burning. 
We shall consider the same mode set than for the Sun, that is $l =0,1$ and 2
and $12\leq n \leq 27$. To obtain estimates of the
errors in these functions we assume
normally distributed frequency errors with $\sigma = 0.5 \, \muHz$.

\begin{table}
\caption{Stellar models considered. $X_c$ is the central hydrogen abundance}
\label{mod}
\begin{center}
\begin{tabular} {ccccc}
\hline
model & $M/M_{\odot}$ & $\Xc$ & $ L/L_{\odot} $  & $\Teff$ \\
\hline
1 & 0.85 & 0.6928 & 0.329 & 5038 \\
2 & 0.85 & 0.0003 & 0.891 & 5466 \\
3 & 1.00 & 0.6928 & 0.718 & 5628 \\
4 & 1.00 & 0.1004 & 1.238 & 5813 \\
5 & 1.30 & 0.6928 & 2.572 & 6509 \\
6 & 1.30 & 0.0516 & 3.695 & 6142 \\
7 & 1.70 & 0.6928 & 8.512 & 8071 \\
8 & 1.70 & 0.0479 &10.994 & 6619 \\
\hline
\end{tabular}
\end{center}
\end{table}

\begin{figure}
 \centerline{\psfig{figure=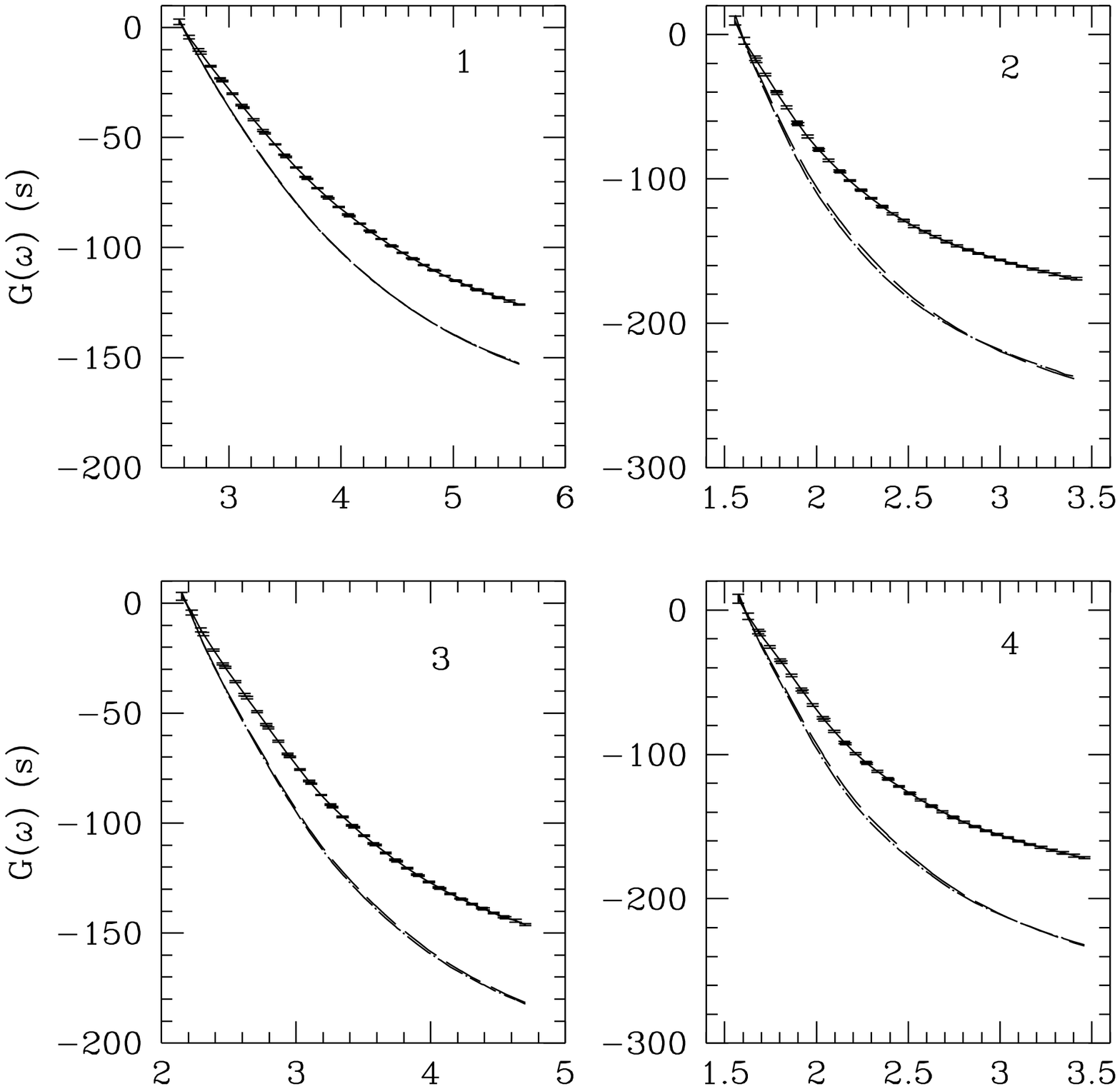,width=.45\textwidth}}
 \centerline{\psfig{figure=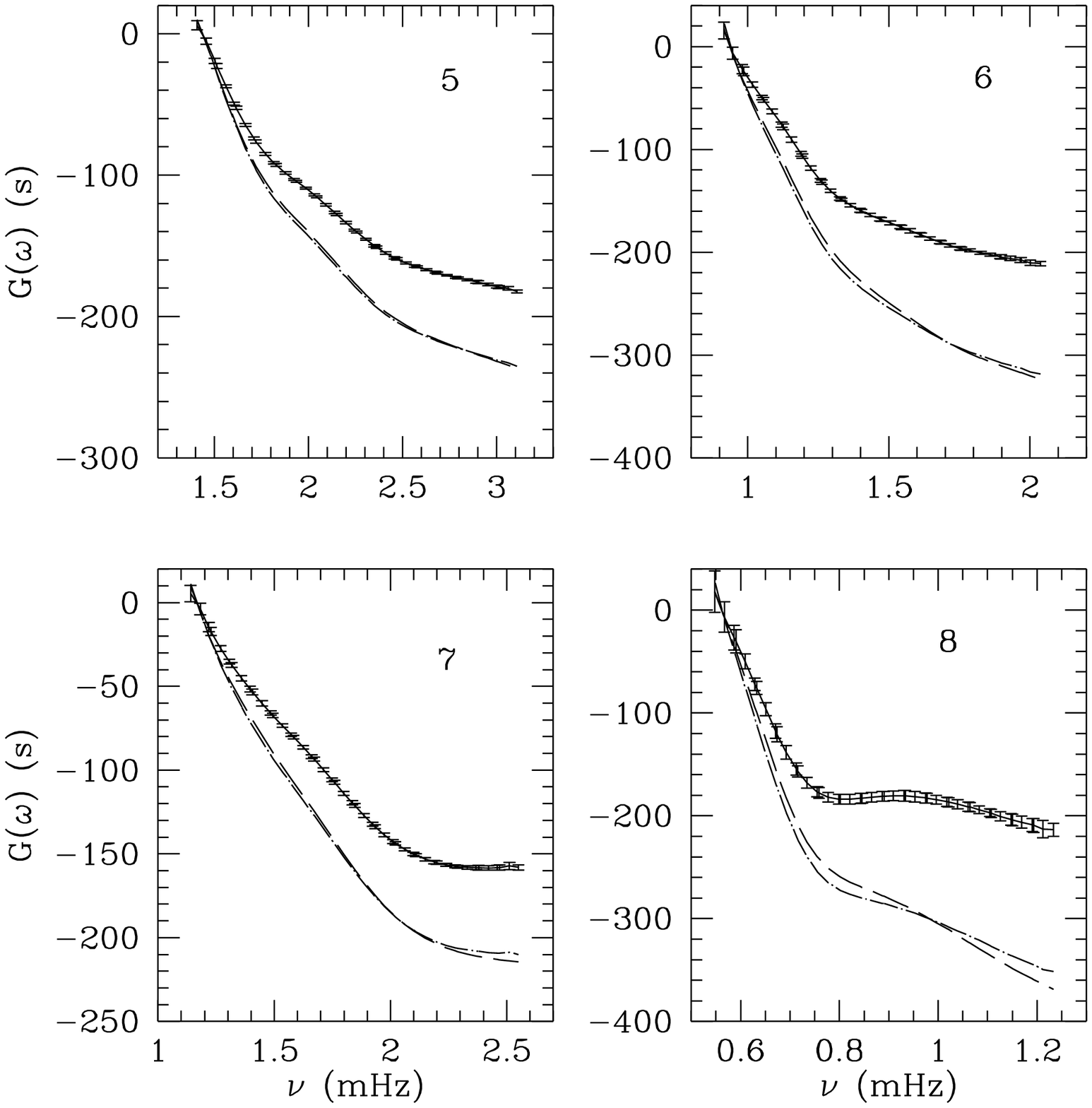,width=.45\textwidth}}
 \caption
{The dashed line is $\Gas$, the continuous line with error bars is
$G$ and the dot-dashed line $G+d/\omega ^2$ for the model of 
Table~\protect\ref{mod} indicated in each panel.
To all the functions we have added a constant so that they match at the 
lowest frequency.}
 \label{fig6}
\end{figure}

Fig.~\ref{fig6} show $\Gas$ -- dashed line -- 
and $G(\omega )$ with the 
corresponding errors -- continuous line -- for all the models considered.
As for the solar case, most of the differences between $\Gas$ and $G$
can be represented as a function of the form $d/\omega^2$, 
as shown in the figures
(the dot-dashed lines correspond to $G+d/\omega^2$). We have also 
computed $G(\omega )$ in the Cowling approximation. Although for clarity 
we do not show it here, we note that in most of the cases 
$\Gas  \simeq \Gcow$, and hence the same comments apply as for the Sun.

However, for the models at the end of hydrogen burning, and in 
particular for the most massive ones, the agreement between $\Gas$ and
$\Gcow$ is worse,
hence indicating that other simplifications leading to 
equation~(\ref{as2}) are not so good for these models.
In fact, as the stars evolve, the buoyancy frequency increases in the core.
This effect is more significant for models with shrinking convective cores,
that is, for relatively massive stars.
As a result, the term in $N^2$ in equation (1) must be taken into
account, and hence the simple asymptotic expression in equation (2)
is no longer adequate.
Furthermore, the buoyancy frequency introduces additional
effects in the correction for the perturbation to the
gravitational potential.
[For more complete discussions, see for example
Vorontsov (1991); Gough (1993); Roxburgh \& Vorontsov (1994).]
Indeed, as shown in Fig.~\ref{fig6}, 
the differences between $G$ and $\Gas$
for the evolved models are not exactly of the form $d/\omega^2$.
Thus analysis of data with very small errors
requires an asymptotic description which is more accurate than equation (2).

\begin{figure}
 \centerline{\psfig{figure=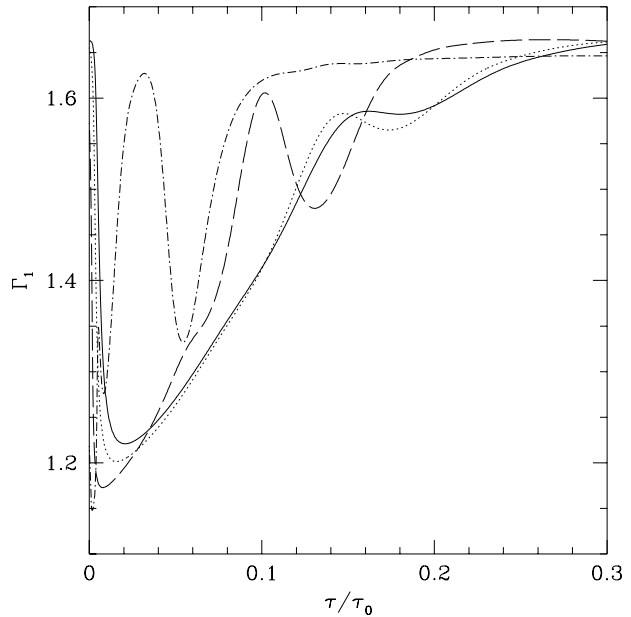,width=.49\textwidth}}
 \caption
{$\Gamma_1$ as function of relative acoustical depth $\tau/\tau_0$
for the ZAMS models in Table~\protect\ref{mod}. The continuous line is for
Model~1, the dotted line for Model~3, the dashed line for Model~5
and the dot-dashed line for Model~7.}
 \label{gamma1}
\end{figure}

To compute the filtered function $\Gf$ we first look for the value of $p$ 
in equation~(\ref{gf}). 
The goal of the analysis is to isolate the signature of the
second helium ionization, as was done in the solar case in Fig.~\ref{fig5}.
Thus the optimal value of $p$ must reflect the acoustical
depth of the ionization region, as indicated by equation (\ref{h2p}).
The location of the second helium ionization varies in temperature
from about \hbox{$1.2 \times 10^5 \K$} in Model~1 to $4.7 \times 10^4 \K$
in Model~7; 
this variation, combined with the change in the stellar surface temperature,
changes the acoustical depth of the associated variation in $\Gamma_1$.
This is illustrated in Fig.~\ref{gamma1}
for a selection of the models in Table 1;
here $\Gamma_1$ is plotted against the relative acoustical
depth $\tau/\tau_0$, where $\tau_0$ is the acoustical radius of the star,
obtained by integrating from $r = 0$ in equation (\ref{tau}).

To determine the optimal $p$, we calculate the response function 
$H(\tau )$ given by equation~(\ref{rf}) for different values of $p$. 
Here the frequency interval and $\alphaas$ are computed for each model 
in Table~\ref{mod}. Then we choose $p$ so as to keep the
information from the second helium ionization zone, here defined at the 
point where $\Gamma_1$ has a relative minimum. We have found that
for Models 1 to 6 in Table~\ref{mod} a value of $p=6$ is suitable. 
For Model~7 a value of $p=2$ has been chosen, and for Model~8, $p=3$.
This decrease of $p$ with increasing effective temperature
clearly reflects the fact that the second helium ionization zone
moves closer to the surface.

\begin{figure}
 \centerline{\psfig{figure=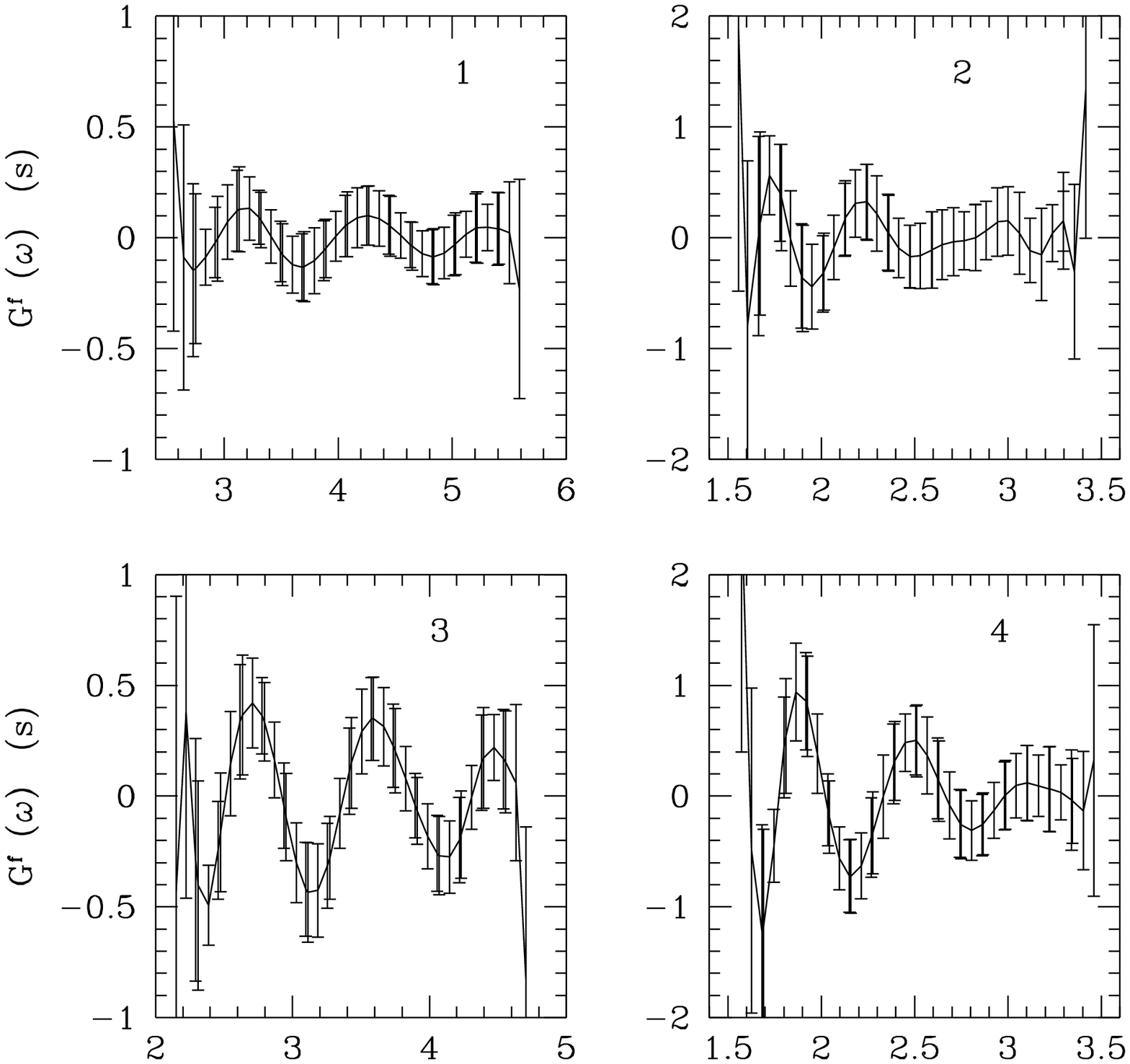,width=.45\textwidth}}
 \centerline{\psfig{figure=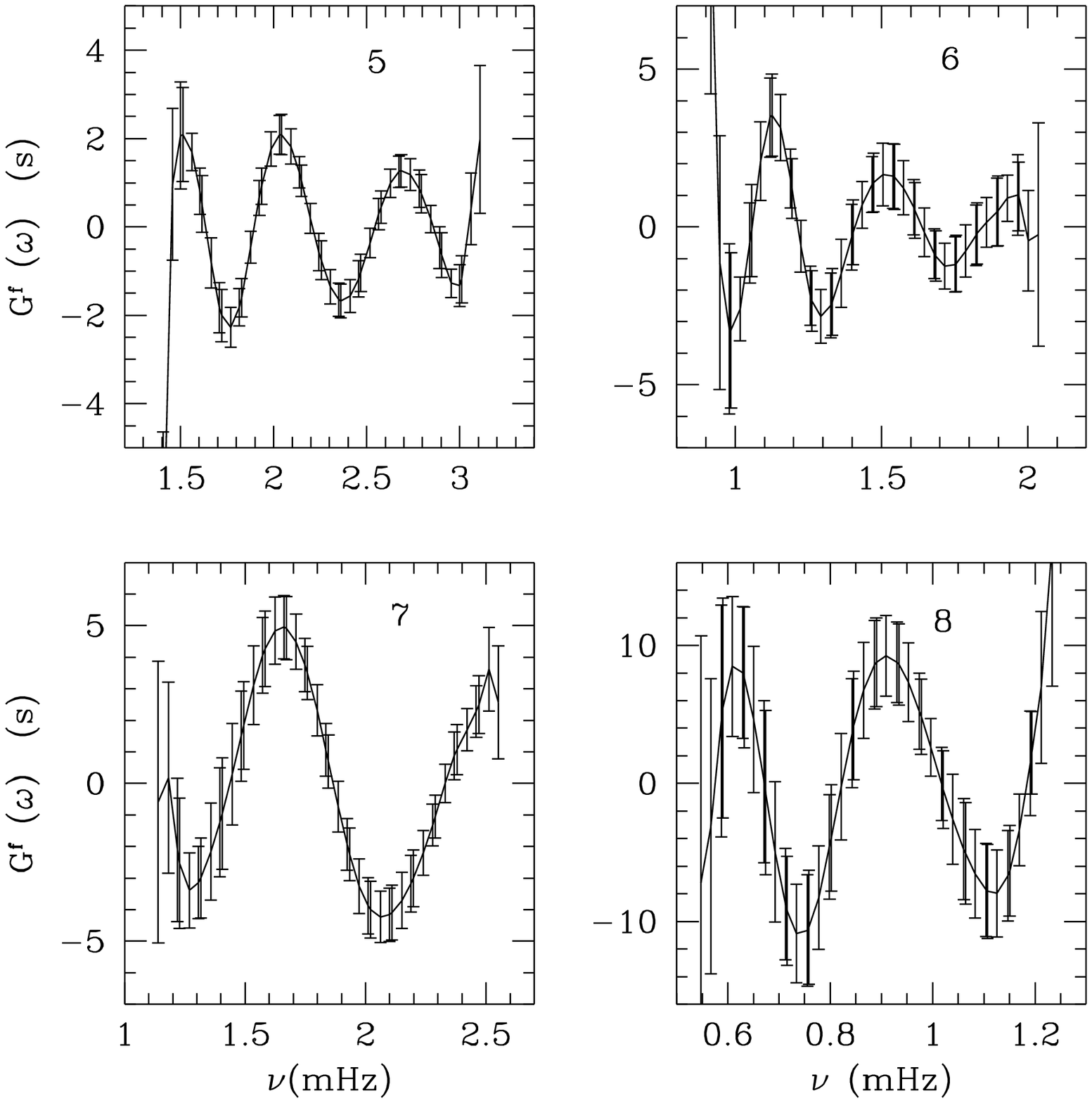,width=.45\textwidth}}
 \caption
{$\Gf(\omega )$ for the model in Table~\protect\ref{mod} 
indicated in each panel.}
 \label{fig8}
\end{figure}

Fig.~\ref{fig8} shows
the functions $\Gf(\omega )$ for the stellar models.
We note that for all the models considered the filtered function $\Gf$ 
agrees within the  errors with the asymptotic $\Gfas$ 
(which, for clarity, we do not show).
As can be seen, for the models with $M=0.85$ $M_{\odot}$, it is not 
possible to detect the signature of the HeII zone as given by $\Gf$ even
with frequency errors as small as $0.5 \muHz$. 
However, for the rest of the models this frequency 
accuracy is sufficient. Also,
from the figure it follows that $\Gf(\omega )$ has a larger amplitude 
for larger masses or later evolution stages,
due to the increased strength of the He-II feature in
$\Gamma_1$ (cf. Fig.~\ref{gamma1}; see also Lopes et al. 1997).
It can also be seen that in the fixed range of radial orders considered,
$\Gf$ contains fewer periods of the signal for $M=1.7$ $M_{\odot}$ 
than for the other masses.
Indeed, it follows from equation (\ref{tas}),
with $\Delta \nu \simeq (2 \tau_0)^{-1}$, that in equation (\ref{h2p})
$2 \omega \tau \simeq 2 \pi n \tau/\tau_0$.
Thus for a fixed range in $n$, such as we have considered here,
the signal arising from the second helium ionization zone 
generally oscillates
less rapidly with decreasing relative depth $\tau/\tau_0$.

\subsection{Changes to the physics and parameters of the models}

We have considered several changes to the input physics and
stellar parameters in order to analyse the sensitivity of the phase function
to such quantities. Since, once the $\pp$-mode frequencies of a star are 
measured, the mean density can be accurately determined, 
it is interesting to compare models with the
same mean density, considering the effects of each modification separately.
To do so, we have computed
envelope models which have boundary conditions only at the surface,
considering models with the same surface parameters as
those given in Table~\ref{mod}. For each 
of these eight models we have considered 
the modifications summarized in Table~\ref{changes}. Here $\alphac$ 
is the mixing length parameter, $X$ the hydrogen abundance, $\kappa$ 
the opacity, $r_c$ the radius at the base of the convection zone 
and EOS refers to
the equation of state; all the models use the EFF 
formulation (Eggleton, Faulkner \& Flannery 1973),
except for the model corresponding
to change 6 which uses the CEFF equation of state 
(see Christensen-Dalsgaard \& D\"appen 1992).
For envelope models it is not possible to compute low-degree $\pp$-mode 
frequencies, and hence we cannot obtain $G(\omega )$ from a fit to 
equation~(\ref{as2}). However, as remarked previously, the
differences in the function $G$ for the models corresponding to the small 
modifications considered in Table~\ref{changes} 
can be expected to be very similar
to those in $\Gas$, which can be computed for envelope models.
	
\begin{table}
\caption{Modifications to the envelope models. } 
\label{changes}
\begin{center}
\begin{tabular} {cccc}
\hline
& parameter & magnitude & notes  \\
\hline
1 & $\Teff$  & +100 K &  \\
2 & $\alphac$ & -0.2 & \\
3 & $X$ & -0.01 & at constant $Z$ \\
4 & $\log_{10}\kappa$ & $-0.2$ in atmosphere & at fixed $r_c$ \\
5 & $Z$ & +0.005 & at constant $X$ \\
6 & EOS & CEFF & \\
\hline
\end{tabular}
\end{center}
\end{table}

\begin{figure}
 \centerline{\psfig{figure=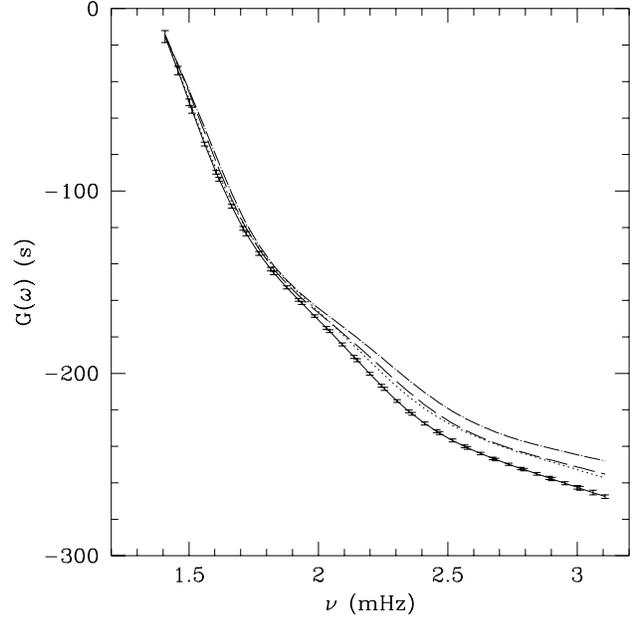,width=.49\textwidth}}
 \caption
{The continuous line is $\Gas(\omega )$ 
for an envelope model similar to Model~5 in Table~\protect\ref{mod}. 
The dashed line corresponds to the change in $\Teff$, the 
dot-dashed line to the change in the mixing-length parameter $\alphac$,
and the dotted line to the change in the atmospheric opacity,
for the changes listed in Table 2.
The functions have been shifted by constants to match at a given frequency.}
 \label{fig10}
\end{figure}

As an illustrative case, we shall show the results for just one case,
an envelope model with the same global and surface parameters as
Model~5 in Table~\ref{mod}. 
In Fig.~\ref{fig10} we show $\Gas(\omega )$ for several
modifications. 
The continuous line is for the
reference model with errors corresponding to frequency errors of 
$0.5 \muHz$. The dashed line corresponds to the change in $\Teff$, the 
dot-dashed line to the change in the mixing-length parameter $\alphac$,
and the dotted line to the change in the atmospheric opacity. 
The functions $\Gas$ for the changes in $X$, $Z$ and the equation of state 
are not shown because they differ from $\Gas$ for the reference model 
by less than the errors. This could be expected since these modifications
do not change significantly the 
structure of the uppermost layers, contrary to those shown in the figure.
Similar results are found for the other models in Table~\ref{mod}.

{}From Fig.~\ref{fig10}, it follows that the function $G$ computed from the 
observations can be used to constrain the structure of the 
uppermost layers. However, it is also clear that
it is not possible to isolate one uncertainty from the rest,
at least with an analysis of a single star. 

\begin{figure}
 \centerline{\psfig{figure=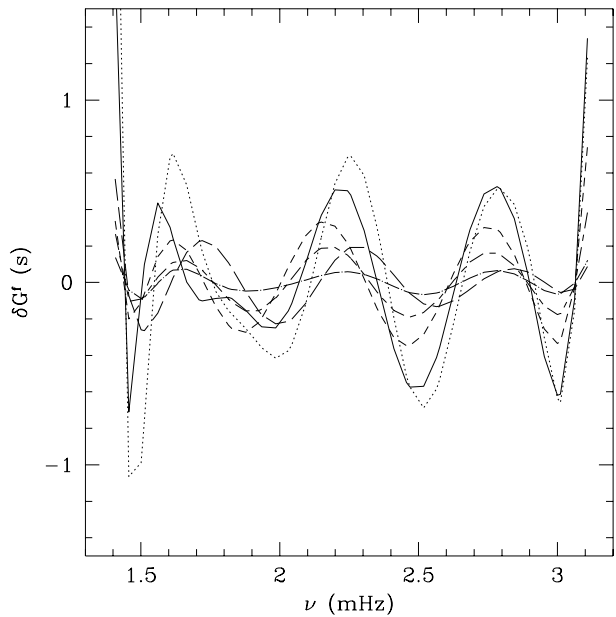,width=.49\textwidth}}
 \caption
{Differences in the filtered phase-function $\Gfas$ between envelope
models with the modifications indicated in Table~\protect\ref{changes} 
and the reference model
(corresponding to Model~5 in Table~\protect\ref{mod}) for: change in 
$\Teff$ (continuous line), 
change in the mixing length parameter $\alphac$ (dotted line),  
change in the atmospheric opacities (long dashed line), 
change in $X$ (dot-dashed line), 
change in $Z$ (short-long dashed line) 
and change in the equation of state (short dashed line).}
 \label{fig11}
\end{figure}

In Fig.~\ref{fig11}, we show the differences $\delta \Gfas (\omega )$ 
between the models with the changes indicated in Table~\ref{changes} 
and the reference
model (an envelope model corresponding to Model~5 in Table~\ref{mod}). 
As expected, $\delta \Gf$ is significant for the change
in the equation of state or the envelope abundances, 
as a result of the corresponding
changes in $\Gamma_1$ in the second helium ionization zone;
however, it is also important for modifications in the atmospheric
opacities or the mixing-length parameter $\alphac$ because these 
modifications change,
for instance, the depth of the second helium ionization layer. This result
is roughly similar to that found for the Sun 
(see P\'erez Hern\'andez \& \jcd \ 1994b), although the relative importance
of each modification is different.

For clarity we have not shown the errors in Fig.~\ref{fig11}
but we note here that for 
frequency errors of $0.5 \muHz$, the mean error in $\Gf$ is of 0.42 s and for
$0.1 \muHz$, the mean error in $\Gf$ is 0.08 s.  
Hence, very accurate frequency determinations are required
in order to obtain information from this function, in particular 
concerning the envelope abundances or the equation of state.

\section{Conclusions}

By analysing low-degree ($l \le 2$) $\pp$-mode frequencies for the Sun we have
shown that it is possible to fit an asymptotic expression which allows to
separate the contribution of the upper layers as given by a function of
frequency $G(\omega )$ from that of the interior. Moreover, it is possible to 
obtain a function $\Gf(\omega )$ that depends mainly on the layers around 
the second helium ionization zone. By applying the same technique to
$\pp$-mode frequencies of stellar models, we have found that the 
same separation is possible for main-sequence stars with masses between 
0.85 $M_{\odot}$ and 1.7 $M_{\odot}$. 

Of course, the main goal is to 
get information about stellar structure from these functions.
When considering observed solar frequencies of low-degree modes,
we have found that the function $G(\omega )$ is determined
with sufficient accuracy
to impose constraints on the structure of the uppermost layers 
of the Sun. Assuming frequency errors of $0.5 \muHz$ we found that the same
can be true for other stars. 
Although different changes to the physics and
parameters of the models
can lead to very similar $G(\omega )$, it might be possible to 
separate the effects of different uncertainties if several stars,
belonging to the same cluster, are analysed;
in this case it may be assumed, for example, that the stars share the
same initial composition.

In principle, analysis of the function $\Gf$ could provide more
significant constraints on the stellar models because it depends on
the structure of the layers around the second helium ionization zone, 
where the physics is better understood than in the uppermost layers.
However, very small errors are
needed in order to obtain useful information from it. 
For instance, in the solar case,
if only low-degree modes are used, frequency errors 
as small as those achieved by GOLF are required in order to 
determine the helium abundance in the solar envelope.
Similar results are found for stars in the main sequence,
for which errors substantially below $0.5 \muHz$ are needed in 
order to impose significant constraints on the stellar models, for 
instance for the helium abundance or the equation of state. 
It is encouraging, therefore, that the COROT mission (Catala et al. 1995)
aims at determining frequencies with errors as small as $0.1 \muHz$.
We note also that $\Gf$ is more sensitive to modifications 
in the equilibrium structure for stars of mass greater than solar.

\section*{ACKNOWLEDGEMENTS}
This work has been made possible thanks to the financial support from the
Spanish DGICYT under grants ESP90-0969 and PB91-0530,
and from the Danish National Science Foundation
through its establishment of the Theoretical Astrophysics Center.

\label{lastpage}

\end{document}